\documentclass[a4paper,11pt]{article}
\usepackage{amsmath,bm,epsfig}

\newcommand{\e}{\varepsilon}

\input epsf

\def\a{\alpha}

\def\g{\gamma}

\def\o{\omega} 

\def\e{\varepsilon}

\def\k{{\textbf{k}}} 

\def\be{\begin{equation}}
\def\ee{\end{equation}}
\def\bea {\begin{eqnarray}}
\def\eea {\end{eqnarray}}

\def\RA {\ \Rightarrow\ }

\begin{document}

\title{Energy spectra of ensemble of nonlinear capillary waves on a fluid surface}

Elena Kartashova and Alexey Kartashov\\

Johannes Kepler University Linz, Austria

\begin{abstract}
Problem of spectral description of the nonlinear capillary
waves is discussed. Usually, three-wave nonlinear interactions
are considered as a major factor determined the energy spectrum
of such waves. We show that the four-wave interactions should be
taken into account. They lead to formation of power energy
spectrum $k^{-\nu}$ with exponent $\nu = 13/6$ (only one
horizontal coordinate) and $\nu = 3/2$ for two-dimensional
propagation.
\end{abstract}

\maketitle

\section{Introduction}

It is a well known fact that the existence of  internal waves in the ocean can be
established in satellite data\emph{ via}
their interaction with short gravity and
capillary waves,  e.g. \cite{EP84,PT99,Jack04}).

The description of the wind ripple is a very difficult task due to
their strong nonlinearity, breaking effects and wind interaction.
Moreover, even if capillary waves are not bounded and have small amplitudes,
their dynamics is not completely understood and a lot of laboratory
experiments have been recently conducted, for instance \cite{SPX10,A09-1,XiSP10,DBF12} and many others.

Theoretically, first analysis of nonlinear interaction of
capillary waves have been done in pioneer work by Zakharov and
Filonenko (1967). In this paper kinetic equation for 3-wave interactions of capillary waves has been first  written out and its stationary solution in the form of power energy spectrum  found.

Kinetic WTT is based on a number of assumptions, the main of them being as follows: \textbf{I}. weak nonlinearity (nonlinearity is small but non-zero and defined by a small parameter $0< \e \ll 1$); \textbf{II}. randomness of phases (all waves interact with each other stochastically); \textbf{III}. infinite-box limit ($L/\lambda\to \infty,$ where $L$ is the size of the system and $\lambda$ is characteristic wave length); \textbf{IV}. existence of an inertial
interval in the wavenumber space $(k_0, k_1)$, where energy input and dissipation are balanced; \textbf{V}. locality of interactions  in $k$-space (only waves with wavelengths of the same order do interact; \textbf{VI}. interactions are locally isotropic (no dependence on direction); \textbf{VII}. at initial moment energy is distributed among an infinite number of modes.

Under these and other assumptions, wave
 kinetic equations  have stationary solutions in the
form of energy power spectra $E_k \sim k^{-\nu}, \, \nu >0, $   \cite{ZLF}. These spectra are called kinetic spectra or K-spectra.

In the case when dispersion function depends only on one dimensional parameter, say
 the gravity constant $g$ for water surface
gravity waves or surface tension constant $\sigma$ for capillary waves, one can compute $\nu$  using dimensional analysis, without solving the corresponding kinetic equation. E.g. for a direct cascade we have:
\be \label{kz_dir}
\nu = 2 \alpha +d-6 + (5 - 3 \alpha -d)/(N-1),
\ee
where $\alpha$ is defined by the form of dispersion function $\omega \sim k^\alpha$, $d$ is the space dimension of
the system and $N$ is the minimal number of waves constituting a  resonance
interaction.

As it was mentioned above, for kinetic WTT to occur,  a number of assumptions \textbf{I}--\textbf{VII} must hold, some of which are not easily verified in laboratory. However the advantage in this case is that the knowledge of dispersion function in a wave system immediately yields the explicit form of energy distribution over the scales.

On the other hand, if we abandon any one of these assumptions, the form of energy distribution will be changed drastically. For instance, in the standard laboratory set up, narrow frequency band excitation is used. In this case, not a statistically described K-cascade is observed, but a D-cascade which  is formed by a set of distinct modes, \cite{CUP}. The spectrum of the D-cascade can be computed deterministically by the increment chain equation method (ICEM) and has exponential form, \cite{K12a}.

In this Letter we take capillary waves as an illustrative example for showing that the standard  approach ``decay-type dispersion function yields automatically 3-wave kinetic regime" is not universal.

\section{Three-wave interactions of capillary waves}
Three-wave resonance conditions for capillary water waves with dispersion function $\o=\sigma k^{3/2}$ read
\be \label{cap3}
k_1^{3/2}+k_2^{3/2} =k_3^{3/2}, \quad \k_1+\k_2=\k_3
\ee

\emph{Case 1.} Wavevectors  $\k_j \in \mathbf{Z}^d$   have integer coordinates $\forall j=1,2,3$ (e.g. wave interactions in a resonator are regarded) and $d$ is arbitrary. In this case (\ref{cap3}) has no solutions for arbitrary dimension $d$ of the wave vectors, \cite{PHD1}.

\emph{Case 2.} Wavevectors  $\k_j \in \mathbf{R}^1$ have real coordinates and $d=1.$  In this case,
\be \label{cap3-1}
k_1^{3/2}+k_2^{3/2} =(k_1+k_2)^{3/2} \RA k_1=0 \ \ \mbox{or} \ \ k_2=0,
\ee
and one can see immediately that for all \emph{positive} $\k_j$ the right hand side of (\ref{cap3-1}) is always greater than its left hand side if both $\k_j \neq 0.$
If $k_1 = k, \ k_2=c k,$ with some constant $1 \le c \le 10$ (cf. \textbf{V}), absolute resonance width
\bea \label{cap3-2}
\Delta_A=|\o_1+\o_2-\o_3|
= |k^{3/2}+c^{3/2}k^{3/2}- [(c+1)k]^{3/2}|\nonumber \\
= k^{3/2} |1+c^{3/2}-(c+1)^{3/2}|> \frac{3\sqrt{c}}{2}k^{3/2}
\eea
is rapidly growing function of $k$ when $k \rightarrow \infty$ (cf. \textbf{III}). In particular, if $k_1 = k_2=k$, $\Delta_A \approx 0.82 k^{3/2}.$

\emph{Case 3.} Wavevectors  $\k_j \in \mathbf{R}^2 $ have real coordinates, $d=2, $ and all three wavevectors are collinear.
This case can  obviously be reduced to the previous one by an appropriate rotation of coordinate axes.

\emph{Case 4.} Wavevectors  $\k_j \in \mathbf{R}^2$ are real valued and non-collinear. One might argue that if in this case a great amount of \emph{almost collinear} wavevectors form approximate triads with \emph{small resonance width}, we still can expect manifestation of 3-wave kinetic regime in laboratory experiments in the form of power energy spectra $E_{k,3} \sim k^{-7/4}$.  This case has been studied numerically and the results are as follows.

\subsection{Resonance width}
Absolute resonance width $\Delta_A$ explicitly depends on $\k_1$ and considering if it is ``small" or ``large" the value of $k_1$ should, of course, be taken into account. It is intuitively clear that for larger vectors larger resonance width is tolerable, and vice versa. Relative resonance width $\Delta_R$, allowing to distinguish between various wave turbulent regimes, might be introduced in a number of ways, e.g. \cite{K09b, zak4} and others; the problems with introducing a general cumulative function $\Delta_R$ are discussed in \cite{CUP}, Ch.6.

To perform numerical study of solutions of  (\ref{cap3}),  for a pair of two-dimensional wave vectors $k_1=(m_1, \, n_1), \quad k_2=(m_2, \, n_2)$ we define relative resonance width as absolute resonance of proportional pair with norm $1$, understanding by the
norm of a pair of two-dimensional vectors that of the corresponding vector in $\mathbf{R}^4$:
$
\|(k_1, k_2)\| = \sqrt{m_1^2+n_1^2+m_2^2+n_2^2}
$
so that
\bea \label{cap3-rrwn2}
\Delta_R=
|((\widetilde{m}_1^2+\widetilde{n}_1^2)^{3/4}+(\widetilde{m}_2^2+\widetilde{n}_2^2)^{3/4} -
((\widetilde{m}_1+\widetilde{m}_2)^2+(\widetilde{n}_1+\widetilde{n}_2)^2)^{3/4})|
\eea
where $\widetilde{m}_j=m_j/\|(k_1, k_2)\|$ and $\widetilde{n}_j=n_j/\|(k_1, k_2)\|$ with $j=1,2.$

\begin{figure*}
\includegraphics[width=5.5cm,height=5cm]{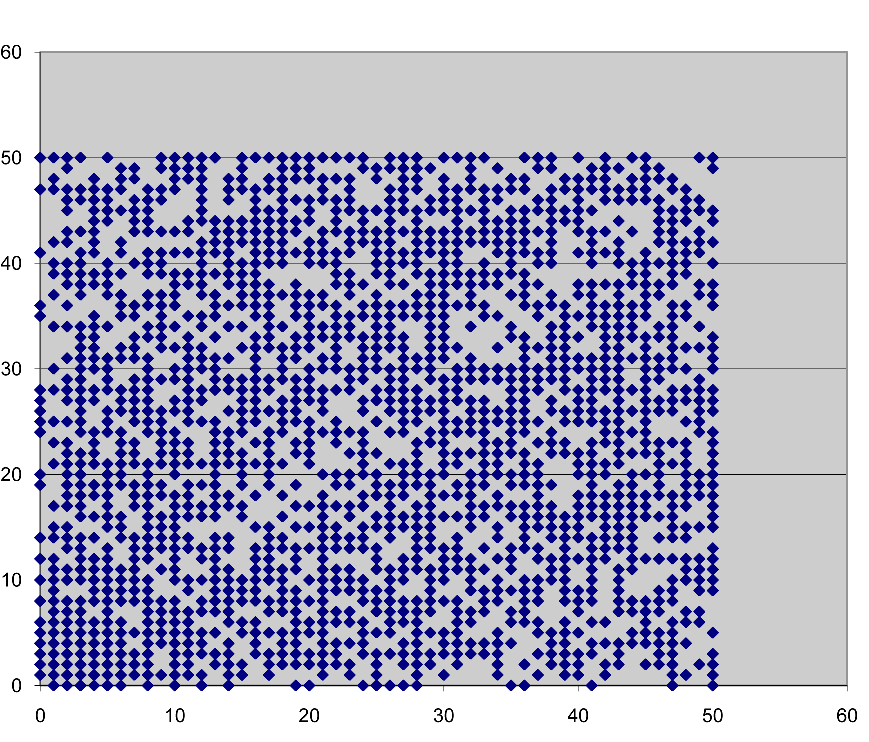}
\includegraphics[width=5.5cm,height=5cm]{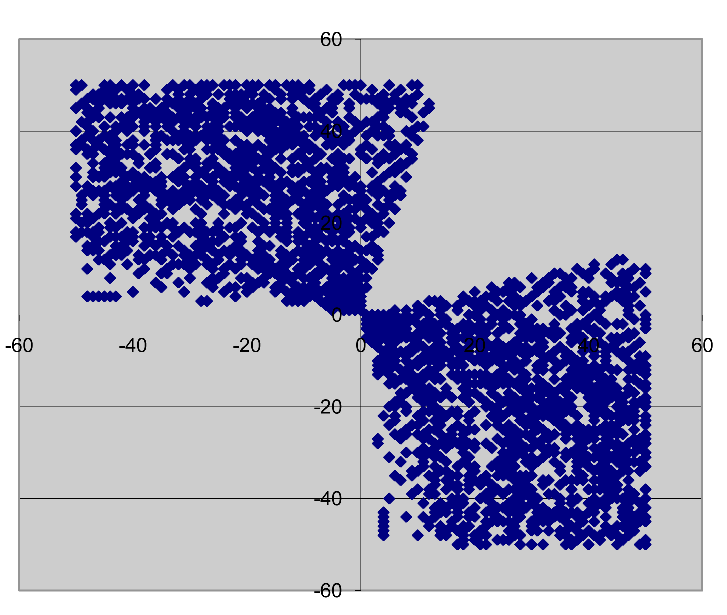}
\includegraphics[width=5.5cm,height=5cm]{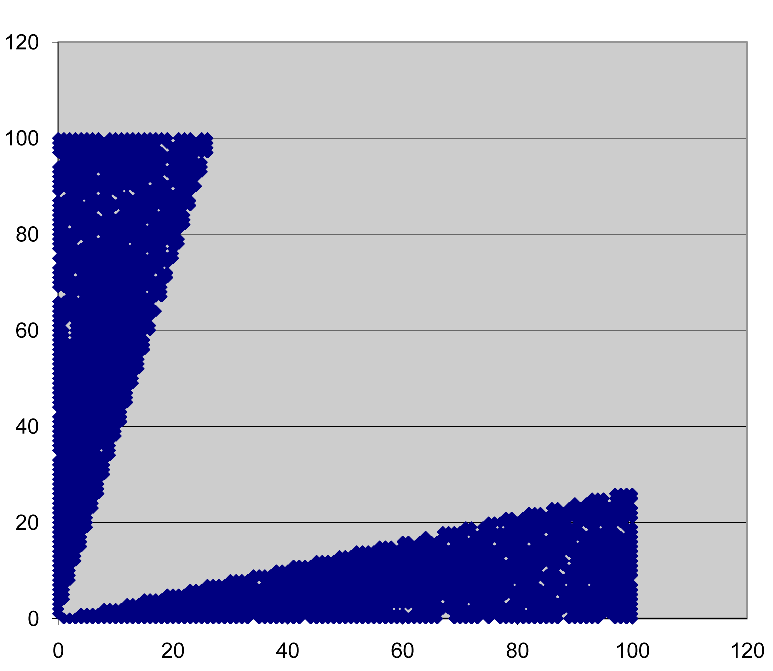}
\caption{\label{WaveNumb} Two-dimensional wavevectors $\k_1, \k_2$ satisfying (\ref{cap3}). \textbf{Left panel}: Wavevectors with non-negative coordinates which interact with vectors with arbitrary (positive or negative) coordinates. Computation domain $-50 \le m_1,n_1 \le 50$. \textbf{Middle panel}: All wavevectors interacting with those shown in the previous panel. Same computation domain.
\textbf{Right panel}: Both wavevectors  have non-negative coordinates. Computation domain $0 \le m_1,n_1 \le 100$. Wavevectors from the lower triangle interact only with vectors from the upper triangle and vice versa.}
\end{figure*}

\begin{figure}
\includegraphics[width=5.2cm,height=5cm]{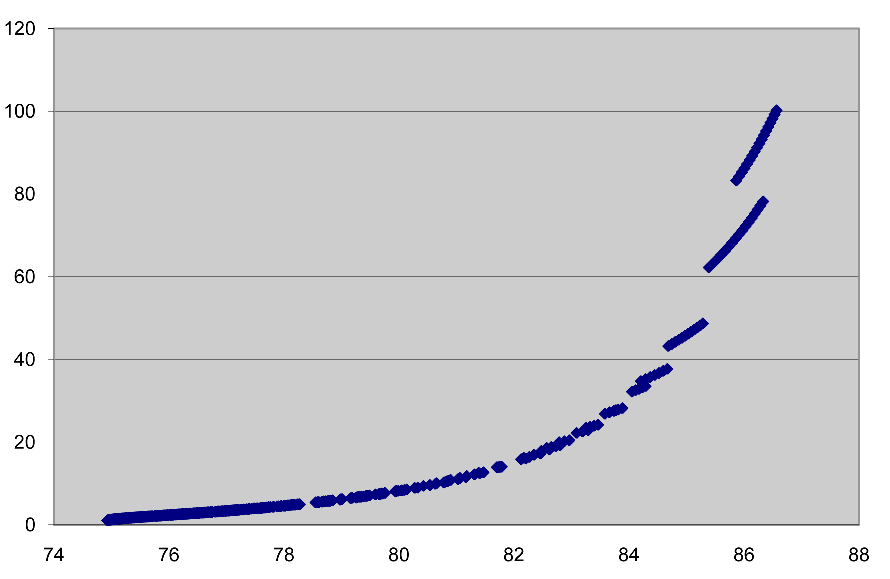}
\includegraphics[width=5.2cm,height=5cm]{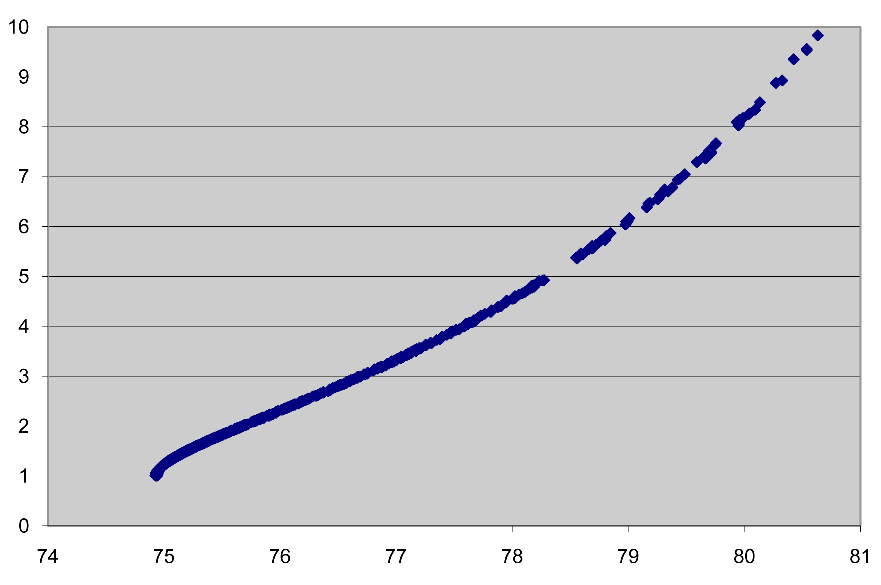}
\caption{\label{Angles}  Dependence of ratio of interacting vectors' norms (longer to shorter) on the angle between vectors. \textbf{Left panel}: Complete picture in computation domain $0 \le m_j,n_j  \le 100$. \textbf{Right panel}: Zoomed presentation of the initial interval (ratio $\le 10$) of the left panel. Axes X and Y denote angles (in grad) and ratios correspondingly.}
\end{figure}

\subsection{Wavenumbers, norms and angles}
 Our first series of numerical simulations served to cast a first glance at distribution of wavevectors satisfying  (\ref{cap3}) in the $k$-space, primarily, if they are distributed evenly over the computation domain or concentrated in some restricted subdomains.
Calculations were performed on $\mathbf{Z}^d$ grid fragments $-50 \, \le \, m_1, \, n_1, \, m_2, \, n_2 \, \le \, 50$ or
$0 \, \le \, m_1, \, n_1, \, m_2, \, n_2 \, \le \, 100$.

Exact resonances (with $\Delta_R=0$) are achieved for pairs $(k_1,0)$ and $(0, k_2)$ only (cf. \emph{Case 1}), while for all other pairs $(k_1, \, k_2)$ approximate interactions may take place.
In the Fig.\ref{WaveNumb}, left panel, all wavevectors $\k_1, \k_2$ with \emph{non-negative coordinates} taking part in approximate interactions are shown, and their distribution appears to be fairly even. However, if one of the wavevectors, say $\k_1,$ has non-negative coordinates, all $\k_2$ interacting with such (middle panel), are distributed in $k$-space quite irregularly, leaving completely empty the third quadrant and the most part of the first quadrant.
Irregularity becomes even more striking if we consider interacting pairs where \emph{both} $\k_1, \k_2$ have non-negative
coordinates (right panel). The most part of the domain consists of wavevectors  \emph{not participating} in interactions, while
interacting vectors are contained in narrow triangles along the axes. Moreover, a simple check shows that wavevectors from the lower triangle interact only with vectors from the upper triangle and vice versa.
are shown in the Fig.\ref{WaveNumb}, left panel.

To characterize the ratios of norms of interacting vectors (cf. \textbf{V}) and angles between them (cf. \textbf{VI}), for each solution we computed the ratio of the vector norms $k_1/k_2$ and the corresponding angle $(\widehat{k_1\,\, k_2})$ (see Fig.\ref{Angles}). It can be seen immediately that solution set is highly anisotropic -- angles between interacting wavevectors all belong to the narrow band  between $75^{\circ}$ and $87^{\circ}$, i.e. interacting wavevectors are almost perpendicular. Norms of the interacting wavevectors can differ by more than 2 orders (Fig.\ref{Angles}, left panel) -- maximal ratio found in our solution set is $k_1/k_2 = 101.8.$ For more than $10\%$ of all the solutions, $k_1/k_2 > 10.$ Restriction of our attention to interactions of wavevectors with norms of the same order makes angle anisotropy even more pronounced (Fig.\ref{Angles}, right panel) -- all angles now lie between $75^{\circ}$ and $81^{\circ},$ i.e. the band width becomes twice smaller. Standard averaging by angles spectra, \cite{zak4}, obviously can not give any reliable information in this case.

\subsection{Resonance curves}
Solution distribution irregularities demonstrated above have an elegant explanation. Indeed, let us notice two simple properties of the resonance set of wavevectors satisfying  (\ref{cap3}):
\begin{itemize}
  \item if a pair $(k_1, \, k_2)$ is a solution, then every $(ck_1, \, ck_2)$ is also a solution for any $c \in \mathbf{R}$
  \item if a pair $(k_1, \, k_2)$ is a solution, then every rotated pair $(\textbf{T}k_1, \, \textbf{T}k_2)$ is also a solution for any $\textbf{T} \in \mathbf{SO}(2,\mathbf{R})$
\end{itemize}
\begin{figure}
\includegraphics[width=5.cm,height=4.5cm]{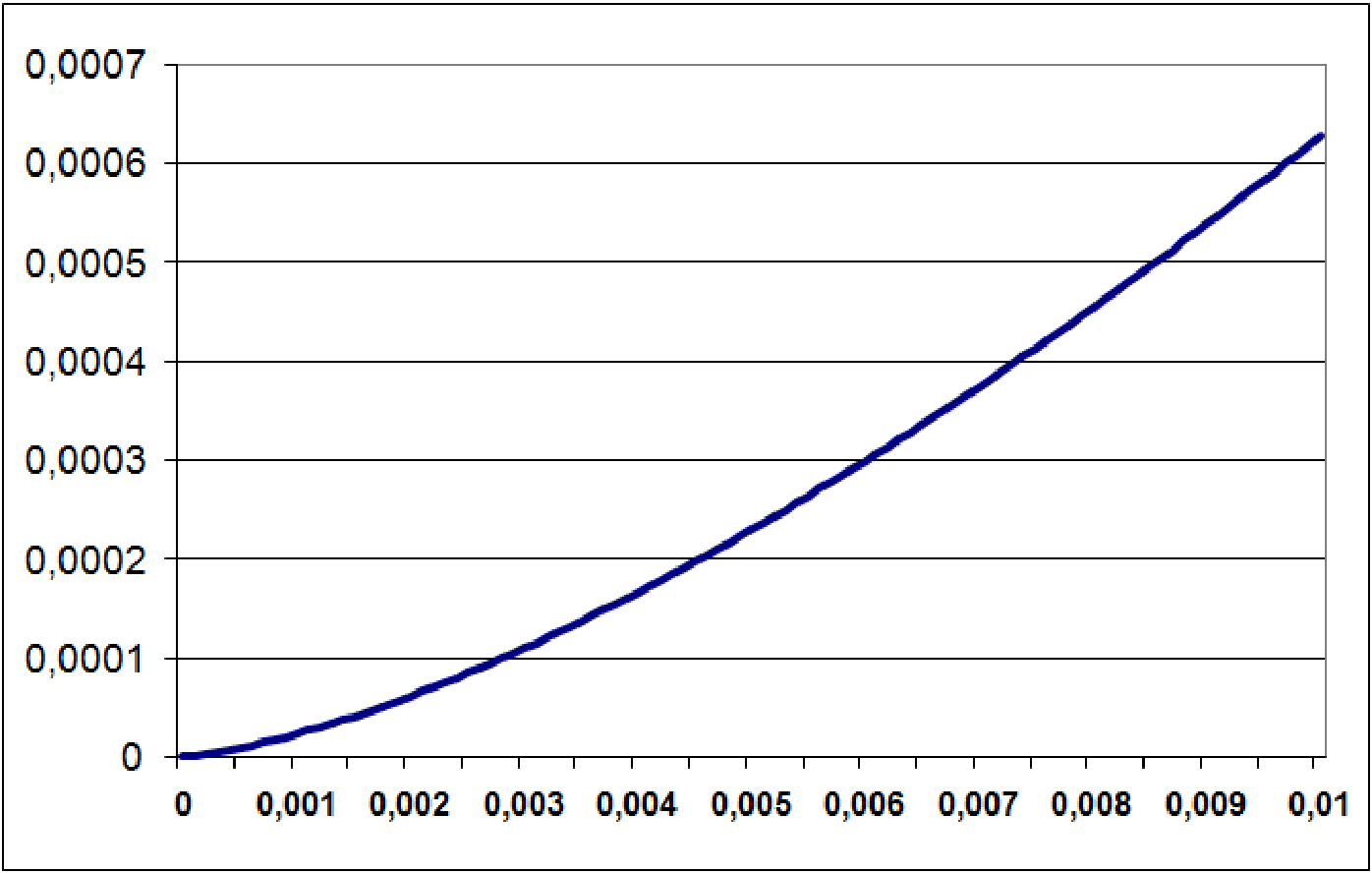}
\includegraphics[width=5.cm,height=4.5cm]{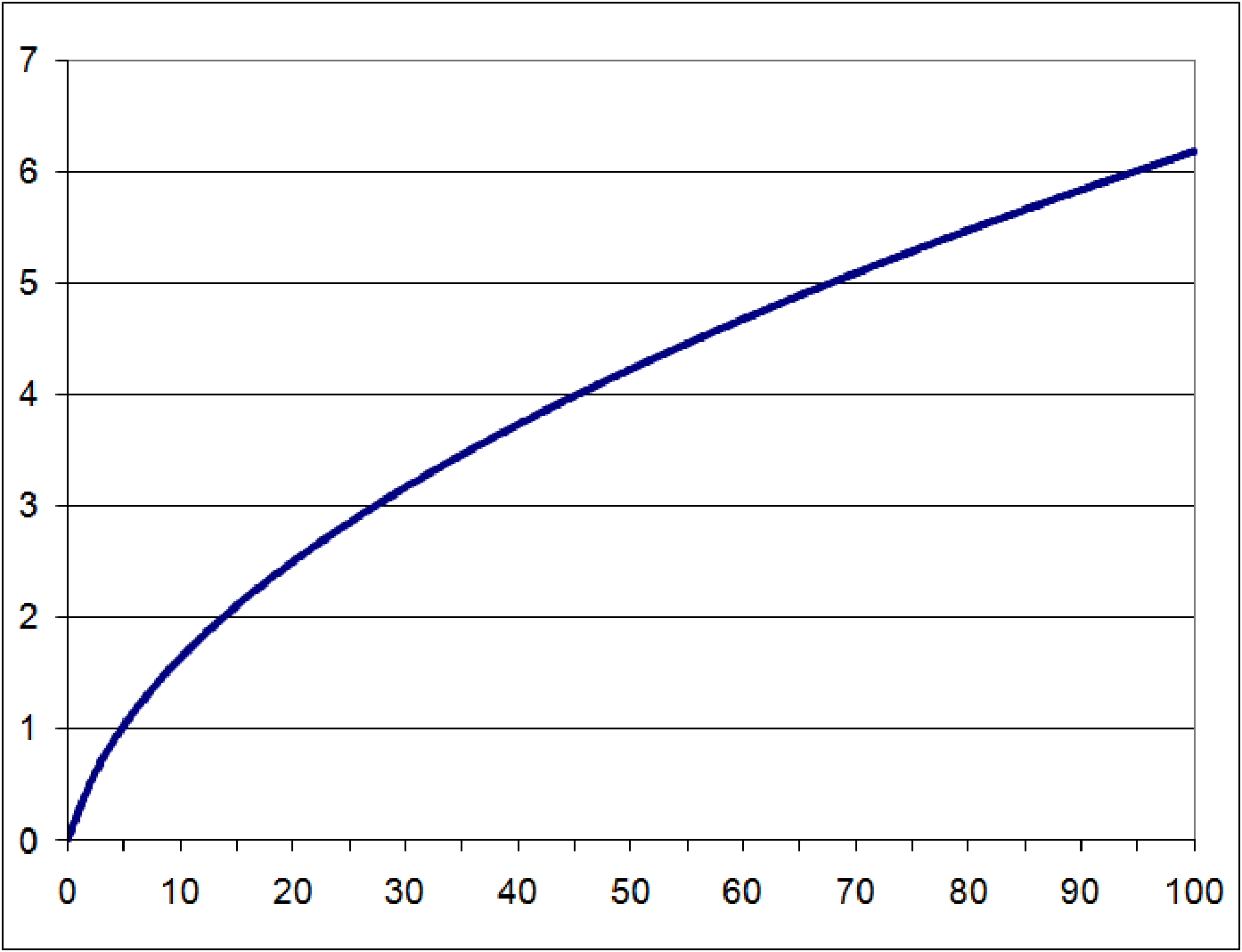}
\caption{\label{ResCurve32} Color online. Resonance curve of vector $(0,1)$ in $k$-space, for dispersion function $\o \sim k^{3/2}$. \textbf{Left panel}: The initial segment of the curve: $m \ll 1 \RA n \sim m^{3/2}$. \textbf{Right panel}: The overall view of the curve: for $m \gg 1 \RA n \sim m^{1/2}$.}
\end{figure}
\begin{figure}
\includegraphics[width=5.2cm,height=3.2cm]{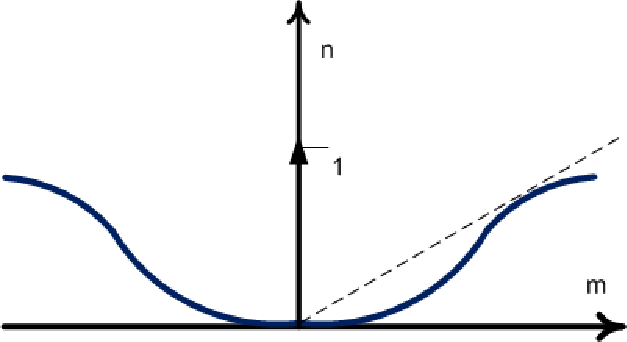}
\includegraphics[width=5.2cm,height=3.2cm]{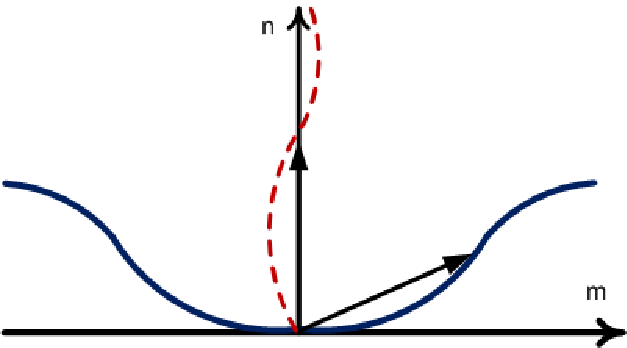}
\caption{\label{ResCurves} Color online. Resonance curves in $k$-space (schematic representation). \textbf{Left panel}: For the vector
$k_1=(0,1)$ all vectors $k_2$ lie on the interaction curve shown. \textbf{Right panel}: Two interacting vectors lie on each
other's resonance curves reciprocally. Resonance curve of the rotated vector is shown by the dashed line.}
\end{figure}

Therefore, it is enough to compute all vectors $k_2$ producing resonant interactions with some given $\mathbf{k_1}$, say $\mathbf{k_1}=(0,1)$ to
obtain a clear view of the whole resonant interaction set. Indeed, all resonance partners of $\mathbf{k_1}=(0,1)$ constitute
a smooth curve shown in Fig. \ref{ResCurve32}. This curve, as a function $n(m)$, starts with a flat region $n \sim m^{3/2}$ (left panel), then becomes steeper and for $m \rightarrow \infty$ has asymptotic $n \sim m^{1/2}$ (right panel). Notice that the two asymptotic regions lie a few magnitudes of 10 apart and can not be illustratively presented in one figure; so we proceed with schematic representation
(Fig. \ref{ResCurves}).

The tangent to the curve drawn from $(0,0)$ gives the $\mathbf{k_2}$ with the minimal angle $(\widehat{\mathbf{k_1}\,\, \mathbf{k_2}}) \sim 74.9°$. We also see that the unit vector can interact both with vectors of arbitrarily small and arbitrarily large norms $k_2$. Notice that both for $k_2 \rightarrow 0$ and $k_2 \rightarrow \infty$ the angle $(\widehat{\mathbf{k_1}\,\, \mathbf{k_2}}) \rightarrow \pi/2$. Now, any vector $\textbf{k} \in \mathbf{R}$ can be produced by stretch and rotation of our unit vector, and its resonance curve is obtained by stretching the curve of the unit vector (with the same coefficient) and rotation (by the same angle).
If two vectors interact resonantly, then each of them lies on the resonance curve of another (Fig. \ref{ResCurves}, right panel).

We  may conclude with confidence that conditions for 3-wave kinetic regime to occur are decidedly violated.

Accordingly, for describing K-spectrum of the system of capillary waves with distributed initial state we have to regard 4-wave resonances, i.e. take $N=4$ in (\ref{kz_dir}).

Indeed, the evidence of strong four-wave coupling in nonlinear capillary waves has been identified in \cite{SPX10} by computing tricoherence as
\be
\tau^2=|\langle F_1F_2F_3F^*_{1+2-3}\rangle|^2/\langle |F_1F_2F_3|^2 \rangle \langle |F_{1+2-3}|^2 \rangle,
\ee
where $F_j$ is the Fourier component of the surface elevation at the frequency $\o_j$. In general, tricoherence $\tau^2 $ can change from 0 (no phase coupling) to 1 (coherent phases); in experiments reported in \cite{SPX10} the level of tricoherence $\tau^2 >0.5$ has been observed.

As K-spectrum relies on the broad excitation and in usual laboratory experiment  we have to deal with narrow frequency band excitation. The standard assumption is that starting with one excited frequency, a distributed state will establish suitable for application of kinetic WTT. The transition from one-mode excitation to the broad excitation is described by dynamic energy cascade formed by the set of distinct modes and can be computed by the increment chain equation method (ICEM), \cite{K12a}. How to apply it for the case of capillary waves is shown in the next section.

\section{Dynamic energy cascade of capillary waves}
The model of the dynamic energy cascade -- D-cascade --  generation has been first proposed in  \cite{CUP}; the physical mechanism underlying formation of a D-cascade is modulation instability.
The phenomenon of modulation instability  has been encountered in various fields and is known under different names --
parametric instability in classical mechanics,  Suhl instability  of spin waves,  Oraevsky-Sagdeev  decay instability
of plasma waves, modulation instability in nonlinear optics, Benjamin-Feir instability in deep water, etc.

Modulation instability is the physical phenomenon which can be described as the decay of a carrier wave  $\o_0$ into two side-bands $\o_1, \, \o_2$:
\bea
\o_1 + \o_2 = 2\o_0, \quad
\vec{k}_1+\vec{k}_2=2\vec{k}_0+ \Theta, \label{ModInst}\\
 \o_1=\o_0 + \Delta \o, \, \o_2=\o_0 - \Delta \o, \,  0<\Delta \o \ll 1. \label{Delta-Omega} \eea
A wave train with initial real amplitude $A$, wavenumber $k=|\vec{k}|$, and frequency $\o$  is modulationally unstable if
\be
0 \le {\Delta \o}/{A k\o} \le \sqrt{2}\label{interv-inst}.
\ee
Eq.(\ref{interv-inst}) described so-called instability interval for the wave systems with a small nonlinearity of order of  $\e\sim 0.1$ to 0.2, first obtained in \cite{BF67}. It is also established for gravity surface waves that the most unstable modes in this interval satisfy the condition
\be \label{BFI-incr}
\Delta \o / A k \o =1.
\ee

The essence of the increment chain equation method is the use of (\ref{BFI-incr}) for computing the frequencies of the cascading modes. At the first step of the D-cascade, excited wave  with frequency $\o_0$ is regarded as the carrier mode. The distance to the next cascading mode $\Delta \o=|\o_0 - \o_1|$ with frequency $\o_0$ is chosen in such a way that  condition (\ref{BFI-incr}) is satisfied; it is called maximum increment condition.

At the next step of the D-cascade, the  mode with frequency $\o_1$ is regarded as a carrier mode for the next step of the D-cascade, and so on.  This procedure can easily be written out as a recursive relation
between neighboring cascading modes:
\be \label{1}
\sqrt{p}_n A_n=  A(\o_n \pm \o_n A_n k_n)
\ee
Here notation $p_n$ is chosen for the fraction of  energy transported from the cascading mode $A_n$ to the cascading mode $A_{n+1}$, i.e. $A_{n+1}=\sqrt{p_n}A_n$. The Eq.(\ref{1}) describes two chain equations: one chain equation with "+" for direct D-cascade with $\o_n< \o_{n+1}$ and another chain equation with "-" for inverse D-cascade with $\o_n> \o_{n+1}$. All computations below are given for direct D-cascade; computations for the inverse cascade are quite similar; they are omitted.

Theoretically $p_n=p_n(A_0, \o_0, n)$  is a function of the excitation parameters $A_0, \o_0$ and the step $n$. However, as in a lot of experiments it is established that $p_n$ depends only on the excitation parameters and does not depend on the step $n$, all the formulae below are given for this case. Accordingly, notation $p$ is used instead of the notation  $p_n$.
This means that $A_{n+1}=\sqrt{p}A_n=p^{n/2}A_0$ and as energy $E_n \sim A_n^2$ it follows $E_n \sim p^n A_0^2$, i.e. energy spectrum of the D-cascade has exponential form as in experimental data for capillary waves, e.g. \cite{XiSP10,DBF12}.

Taking Taylor expansion of the RHS of the chain equation and regarding only two first terms of the resulting infinite series, one can derive a very simple ordinary differential equation describing stationary amplitudes of the cascading modes satisfying (\ref{BFI-incr}):
\bea
\sqrt{p} A_n \approx A_n + A_n^{'}\o_n A_n k_n
\RA A_n^{'} = \frac{\sqrt{p}-1}{\o_n k_n } \RA  \label{2terms}  \\
A(\o_n) =  (\sqrt{p}-1) \int \frac{d \o_n}{\o_n k_n}+C(\o_0, A_0) \qquad \label{A-gen}
\eea
 where $\o_0, A_0$ are excitation parameters.

 The maximum increment condition for the weakly nonlinear capillary waves with $\e \sim 0.1\div 0.2$ differs from (\ref{BFI-incr}) by the constant coefficient 1/24:
\bea
(\Delta \o) /\Big(\frac{1}{24} \o A k \Big)=1, \label{cap-small}
\eea
as was first shown in \cite{Hog85}. As for capillary waves $\o(k) \sim k^{3/2}$, one gets easily e.g. for  direct D-cascade that
 \bea
 (\sqrt{p} -1) \approx \frac{1}{24} A_n^{'}\o_n^{5/3}  \RA \\
E(\o_n)^{(Dir)} \sim \Big[ \frac{(1-\sqrt{p})}{16} \o_n^{-2/3}+C^{(Dir)} \Big]^2,\quad
\mbox{where}\qquad C^{(Dir)} = A_0 - \frac{(1-\sqrt{p})}{16} \o_0^{-2/3}. \label{cap-D-small_E}
\eea

\section{K-spectrum \emph{vs} D-spectrum}
\begin{figure}
\begin{center}
\includegraphics[width=6cm,height=5.5cm]{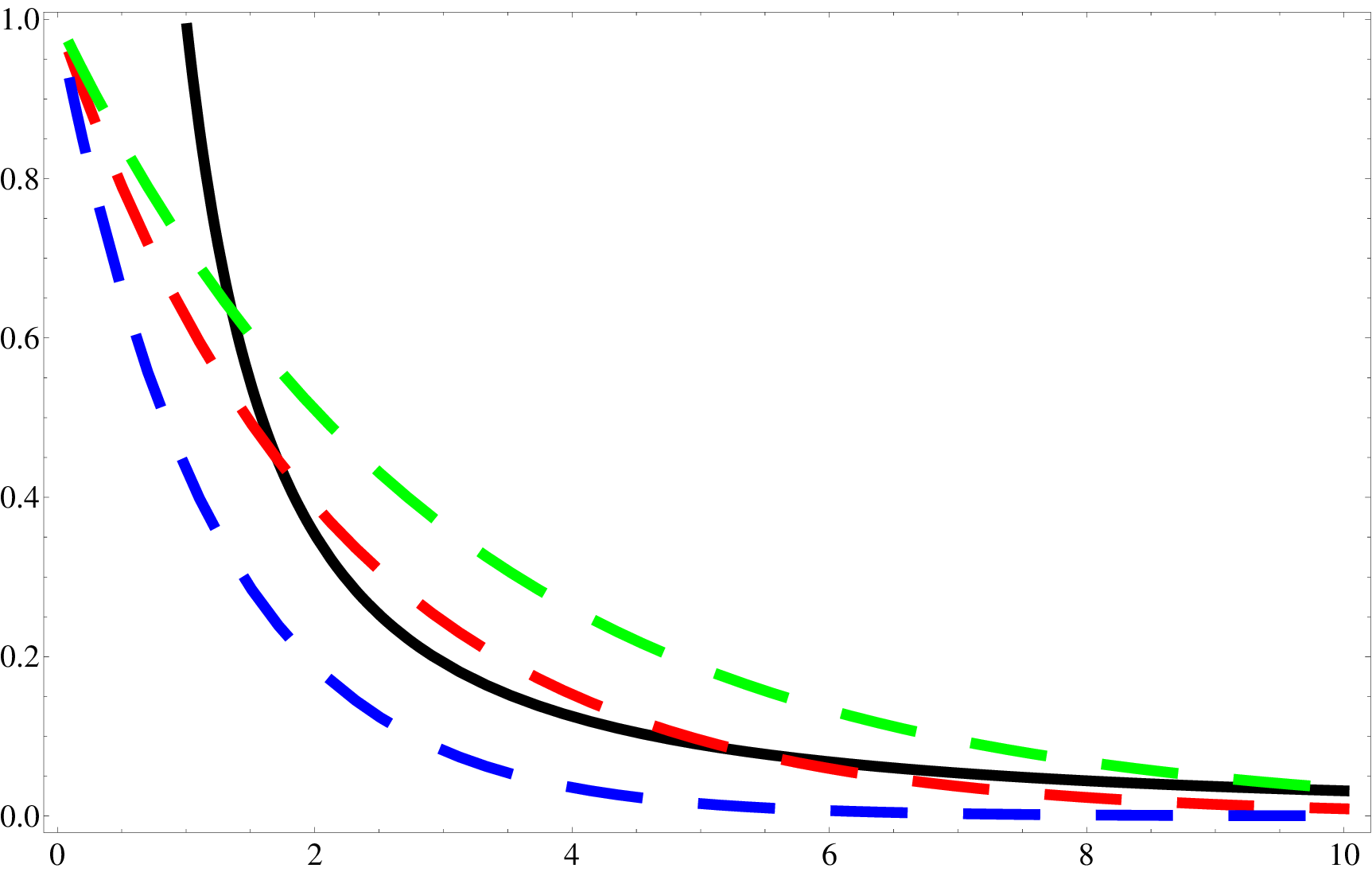}
\includegraphics[width=6cm,height=5.5cm]{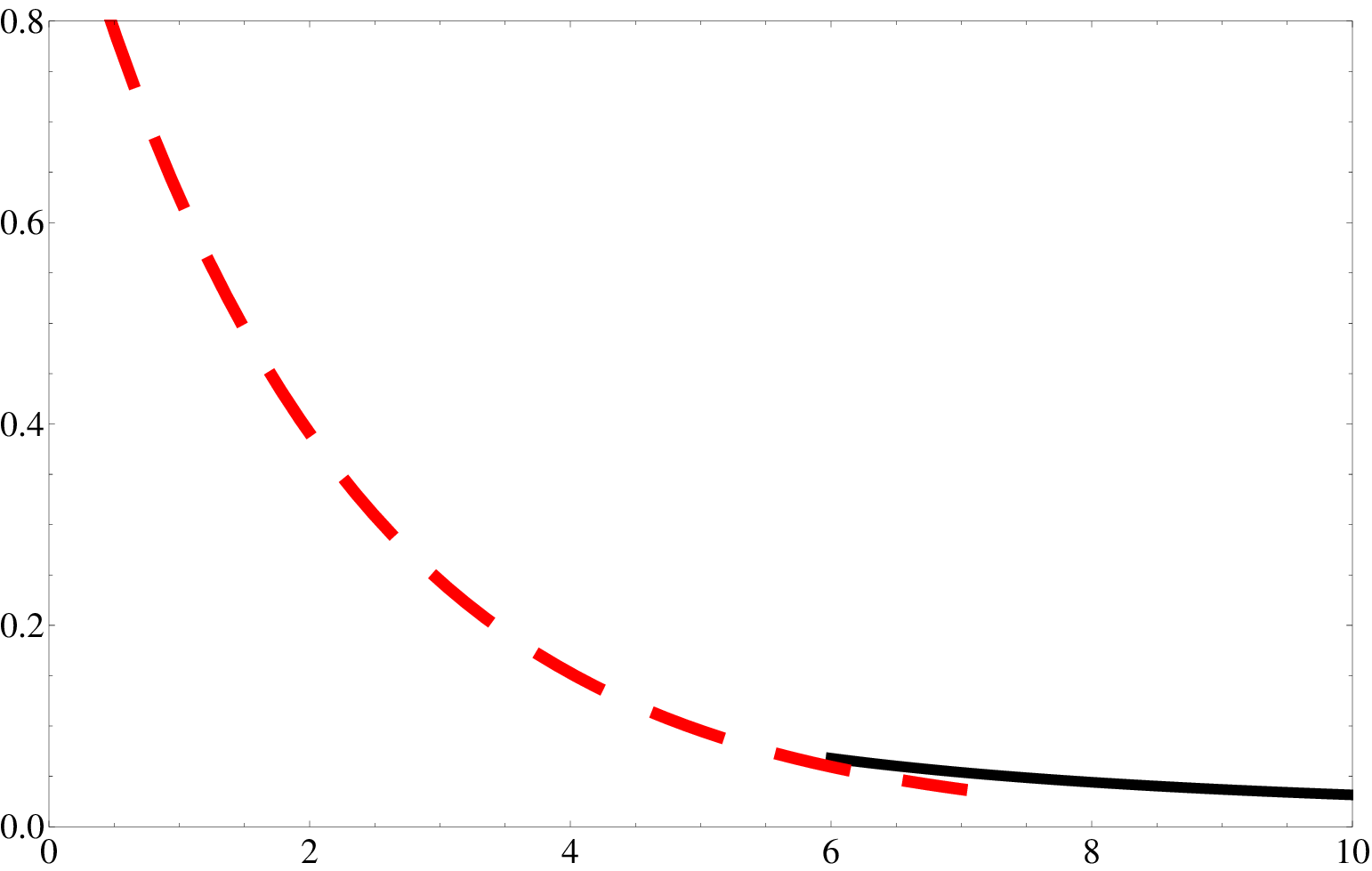}
  \caption{Color online. In both panels, function $x^{-1.5}$ is shown by bold black line. Function $b^{-x}$ is shown by dashed lines of various colors for  $b=1.4; 1.6$ and 2.3}\label{f:cascades}
  \end{center}
\end{figure}

For comparing  energy spectra $E_k$ and $E_n,$ it is convenient
to rewrite $E_n$ as $E_n=b^{-n}E_0$ with $b=1/p, \ b>1.$ Thus we have to compare functions $\g_1 \cdot b^{-x}$ and $\g_2 \cdot x^{-a},$ where the magnitudes of parameters $a,b,\g_1, \g_2$ are defined by the specific wavesystem. As for $a,b>1$
\be
\lim_{x \rightarrow \infty}({x^{a}}/{b^{x}})=0,
\ee
\noindent  $E_k > E_n$ in the long run.
However, for some combinations of parameters and in some finite domains in $k$-space, the opposite relation can take place, $E_k < E_n$;
the spectra $E_k$ and $E_n$ might  be quite close and even coincide for some $k$ (see Fig.\ref{f:cascades}, left panel). Main characteristics allowing to distinguish between kinetic and dynamic cascades which can be easily observed in experimental data are summarized in the Table below.
\begin{center}
\begin{tabular}{|c||c||c|}
\hline
Property & $E_k$ & $E_n$\\
\hline \hline
coherent phases & no & yes\\
\hline
dependence on the excitation parameters & no &yes\\
\hline
local interactions & yes & no\\
\hline
existence of inertial interval & yes & not important\\
\hline
small parameter& $\sim 10^{-2}$ & $\sim 10^{-1}$\\
\hline
\end{tabular}
\end{center}

As formation of the D-cascade is accompanied by the spectrum broadening, at some moment of time
 phases become stochastic, and $s$-wave resonant interactions may appear and kinetic regime may be developed (shown schematically in the Fig.\ref{f:cascades}, right panel).

This scenario seems to be confirmed in laboratory experiments with parametrically excited capillary waves \cite{XiSP10}, the container
shaken at frequencies from 0.5 to 3500 Hz. Energy contained in a zero-frequency band and a dynamic cascade are observed; they contain total energy $E_{tot}$ of the system at lower forcing. Kinetic cascade occurs first at frequencies about 220 Hz and its energy grows (with increase of  the forcing frequency) from $0.01\cdot E_{tot}$ to $0.23\cdot E_{tot}$, while energy contained in the
dynamic cascade decreases from $0.82\cdot E_{tot}$ to $0.46\cdot E_{tot}.$

The understanding of differences between dynamic and kinetic cascades is of the utmost importance for correct interpretation of the experimental observations. Thus, in \cite{A09-1} weak turbulence of capillary waves in Helium has been studied and the formation of a local maximum of the wave-spectrum near a viscous cut-off was observed (under periodic driving force) and correctly attributed to the discrete regime (interactions are non-local).

On the other hand, the authors conclude that "in the inertial range dependence of the peak amplitudes on frequency is described well by a power law function $I_{\o} \sim \o^{-m}$ with the index $m \approx 3.7. $ This is in agreement with the weak turbulence theory which gives the value $m=21/6$" (\cite{A09-1}, p.032001-3).

As 21/6=3.5, the observed and predicted indexes differ by about $6\%$. It would be worth to check phase coherence in this data in order to understand whether this discrepancy is due to the available accuracy of measurements or while in fact a dynamic cascade is observed and not a kinetic one.

As the form of D-cascade and K-cascade can be pretty similar for some parameters of initial excitation, the main characteristic which should checked while estimating the measured data are time scales for the cascade formation as explained in details in \cite{K13-1}.

\section{Conclusions}

In the system of weakly nonlinear capillary waves  two types of energy cascades are theoretically predicted: K-cascade in the systems with distributed initial state and D-cascade in the systems with narrow frequency band excitation.

As we have shown above, a K-cascade among capillary waves can not be formed  by 3-wave resonant interactions; 4-wave resonant interactions should be regarded instead.
 Accordingly, a K-cascade of capillary waves is formed at the time scale $1/\varepsilon^{4}$ with $\varepsilon \sim 10^{-2}$.

 On the other hand, a D-cascade is always formed at the time scale $1/\varepsilon^{2}$ with $\varepsilon \sim 10^{-1}$, i.e. it is formed much faster  than  a K-cascade. For instance, for capillary water waves with the dispersion function $\o^2=\frac{\sigma}{\rho} k^3$, the density $\rho=10^3kg/m^3$ and the coefficient
of surface tension $\sigma=72,75\cdot10^{-3}kg\cdot m/sec^2$ it is easy to compute corresponding characteristic times. Indeed, say for wave length 1 millimeter we have: wave period is 0,0022 sec;
time scale for D-cascade formation is 0,22 seconds and time scale for 4-wave K-cascade is 2200 seconds which is approximately 37 minutes.

Known laboratory experiments with capillary waves confirm the time scale of the D-cascade, e.g. \cite{SPX10,XiSP10,DBF12}. Accordingly, we conclude that energy cascades of capillary waves observed experimentally are D-cascades and not K-cascades.

This fact has also been noticed in numerical simulations \cite{Pu99,PZ00} and was coined by the term "frozen turbulence". It was observed that capillary waves
demonstrate fluxless modes,
"there is virtually no energy absorption associated with
high-wavenumbers damping in this case" (\cite{PZ00}, p.107). This fact has been attributed to the interplay of two facts:  discretization of the numerical scheme and the absence of
 exact 3-wave resonances among capillary waves with \emph{integer wave numbers}, first proven in \cite{PHD1}.

 The main novelty of the present paper is can be formulated as follows. Though there exists infinite many 3-wave resonances among capillary waves with \emph{real wave numbers}, nevertheless they do not form a 3-wave K-cascade while they do not satisfy the basic assumptions of kinetic WTT.

Moreover, speaking very generally, if dispersion function $\o(\k)$ has  decay type, this only means that 3-wave resonance conditions
\be \label{3general}
\o(\k_1)+\o(\k_2) =\o(\k_3), \quad \k_1+\k_2=\k_3
\ee
may have  solutions with real $k_j$, even infinite number of solutions. However, \emph{this does not necessary mean} that these solutions possess the properties \textbf{I}--\textbf{VI}.
In particular, if $\o(\k) \, \sim \, k^\g, \ \g>1, $ then both properties formulated in Sec.2B hold and the geometry of resonances can be outlined in terms of resonance curves similar to those shown in Fig. \ref{ResCurves}.

The results presented in this paper are obtained for an ensemble of free nonlinear
capillary waves formed from initial monochromatic disturbance.
Next step will be an analysis of the ensemble of capillary waves in
present of current induced by the internal waves.

\noindent \textbf{Acknowledgements.} E.K. acknowledges
the support of the Austrian Science Foundation (FWF) under the project
P22943-N18 ``Nonlinear Resonances of Water Waves".

\end{document}